\documentclass[10pt,letterpaper]{article} 

\usepackage{ol2}
\usepackage[draft]{hyperref}
\usepackage{amsmath}

\usepackage{graphicx}
\usepackage{gensymb}

\begin{document}
\title{Full Field Supercritical Angle Fluorescence Microscopy for live cell imaging}


\author{Thomas Barroca,$^1$ Karla Balaa,$^{1,2}$ Julie Delahaye,$^1$ Sandrine L\'{e}v\^{e}que-Fort,$^2$ and Emmanuel Fort$^{1*}$}

\address{
$^1$ Centre d'Imageries Plasmoniques Appliqu\'{e}es, Institut Langevin, ESPCI ParisTech, CNRS UMR 7587, 10 rue Vauquelin, 75 231 Paris Cedex 05,
France.\\ $^2$ Institut des Sciences Mol\'{e}culaires d'Orsay and Centre de photonique Biom\'{e}dicale (CLUPS), CNRS – Univ. Paris-Sud 11, F91405
Orsay cedex (France).\\ $^*$Corresponding author: emmanuel.fort@espci.fr }

\begin{abstract}We introduce a full field fluorescence imaging technique with axial confinement of about 100~nm at the sample/substrate interface.
Contrary to standard surface imaging techniques, this confinement is obtained through emission filtering. This technique is based on supercritical
emission selectivity. It can be implemented on any epifluorescence microscope with a commercial high numerical aperture objective and offers a real
time surface imaging capability. This technique is of particular interest for live cell membrane and adhesion studies. Using HEK cells, we show
that one can observe simultaneously the surface and in-depth cell phenomena.\end{abstract}

\ocis{170.2520, 260.6970, 220.0220, 240.0240.}

\noindent Numerous cell mechanisms, like membrane trafficking and adhesion processes, are located in direct vicinity of the
membrane~\cite{Bretscher1998}. The understanding of such processes is of crucial importance in many biomedical issues. Because of the minute
concentrations of biomolecules involved, fluorescence microscopy is the widely favoured technique to investigate such systems. However, the
diffraction limited axial sectioning is not sufficient to observe membrane processes separately from the inner cell activity.

In this context, Total Internal Reflection Fluorescence (TIRF) configuration has emerged as a powerful surface imaging
technique~\cite{Axelrod2001}. Axial sectioning is obtained by illuminating above the critical angle. It produces an evanescent field which excites
only fluorophores in the direct vicinity of the interface. Excitation confinement enables one to observe minute concentrations of fluorophores at
the interface while reducing the background noise from the inner part of the cell. The development of high numerical aperture immersion oil
objective lenses has popularized this technique in a through-the-objective configuration. The sensitivity of this technique is mainly limited by
the intrinsic light scattering into the cells which eventually results in a loss of excitation light confinement~\cite{Oheim}.

\begin{figure}
\centerline{\includegraphics[width=8.4cm]{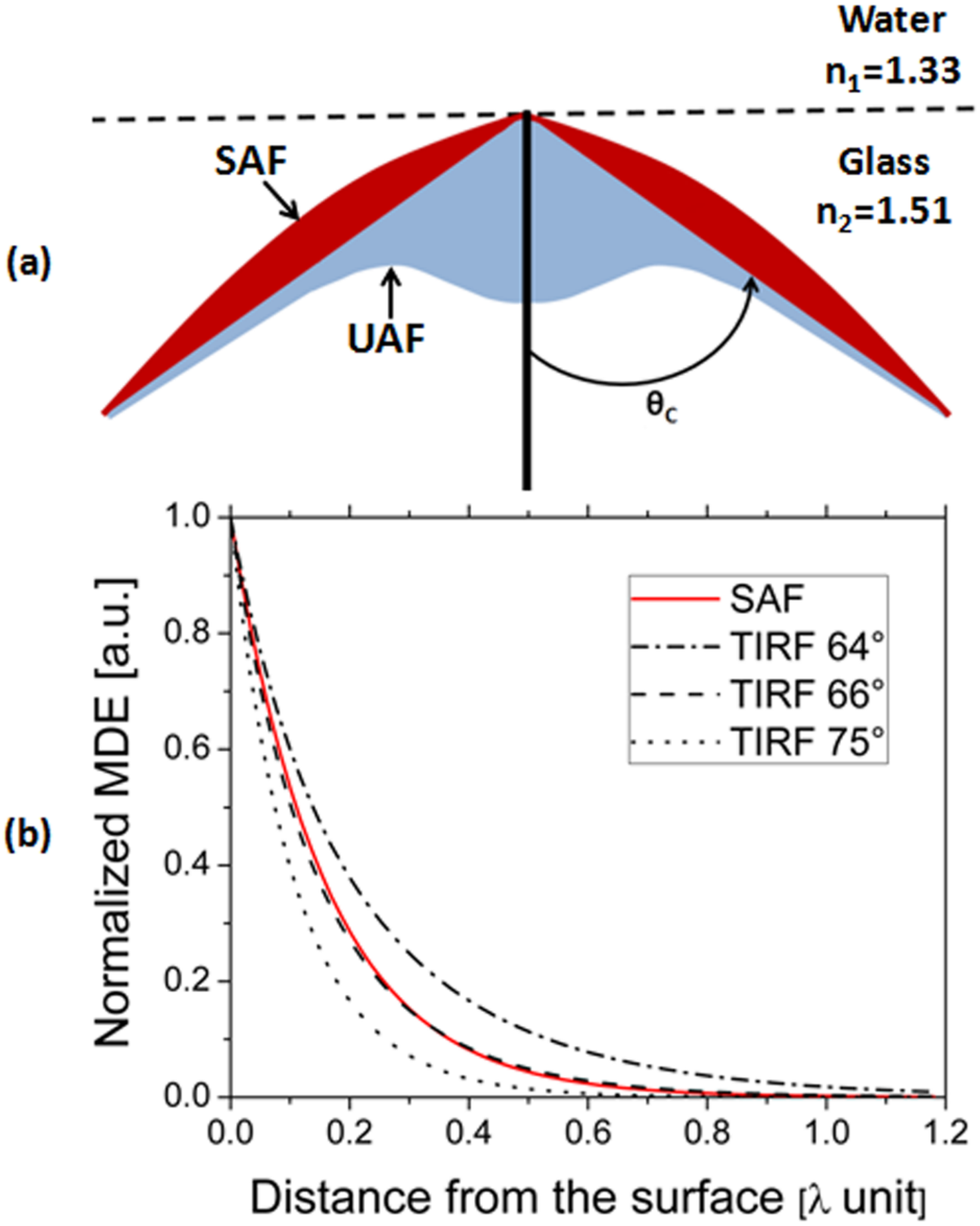}}
\caption{(a): Polar plots of the emission pattern for an isotropically oriented fluorophore positioned at the glass/water interface. (b): Molecular
Detection Efficiency vs fluorophore/interface distance for SAF and various TIRF angles.}
\label{Emission_fluo}
\end{figure}

An alternative approach has been proposed to obtain axial sectioning~\cite{Ruckstuhl2000}. It takes advantage of the fluorophore/interface distance dependent emission.
When a fluorophore in medium~1 (refractive index $n_1$) is placed in the direct vicinity of the interface with a medium~2 (refractive index
$n_2>n_1$), part of the fluorescence is emitted above the critical angle $\theta_{C}=\arcsin(n_1/n_2)$. This supercritical fluorescence is also
called "forbidden light" since Snell-Descartes refraction law does not allow such an emission~\cite{Ruckstuhl2000}. Supercritical Angle Fluorescence
(SAF) originates from the fact that evanescent components of the emission dipole can become propagative in medium~2. Thus, SAF contribution
decreases sharply with the fluorophore/interface distance $d$, contrary to the Undercritical Angle Fluorescence (UAF) components that remain
constant~\cite{Ruckstuhl2004}. In order to compare SAF with TIRF, we study the Molecular Detection Efficiency (MDE) as a function of the
fluorophore/substrate distance $d$, for both the techniques. MDE is given by~\cite{Revue2008}:
\begin{equation}
MDE(d) = \Gamma_{exc}(d) \times QY(d) \times MCE(d)
\end{equation}
where $\Gamma_{exc}$ is the excitation rate, $QY$ is the quantum yield of the fluorophore and $MCE$ is the Molecular Collection Efficiency. For
simplicity, we will assume that the fluorophore has $QY$~=~1. In TIRF, the variation of MDE with $d$ is dominated by the excitation profile
$\Gamma_{exc}(d) \propto \exp(-d/\delta)$ where $\delta$ is the penetration depth of the evanescent field. $\delta$ is function of the wavelength
$\lambda$, the incident angle $\theta_{inc}$ and the refractive indices $n_1$ and $n_2$. In contrast to TIRF, for the SAF technique MDE is dominated by
$MCE(d)$ while $\Gamma_{exc}$ is roughly constant. Figure~\ref{Emission_fluo}(a) shows the emission lobe in glass for a fluorophore placed at the
interface ($d=0$). The SAF emission appears in red and represents about 50\% of the emission in the glass~\cite{Revue2008}. The normalized MDE in
TIRF for different angles $\theta_{inc}$ and in SAF are represented figure~\ref{Emission_fluo}(b). The MDE in TIRF has been obtained multiplying
the (UAF+SAF) $MCE$ by the TIRF excitation profile $\Gamma_{exc}^{TIRF}$. For the SAF, the normalized MDE is simply given by the $MCE$. We
calculate the $MCE$ for UAF and SAF using the vectorial Debye integral model\cite{Wolf1959}. The calculated theoretical penetration depth in SAF is
equal to 129 nm (with $\lambda$~=~593~nm, $n_1$~=~1.33 and $n_2$~=~1.51) which is the penetration depth of TIRF at $66\degree$.

Recently, an implementation of the SAF technique has been proposed using a specifically designed objective lens with a parabolic mirror to collect
supercritical angles~\cite{SAF_Biosensor}. This objective lens has the advantage of simplicity and low cost which is of particular interest in
biosensing applications~\cite{SAF_PRL}. However, since imaging with this objective involves objective/sample scanning, it is relatively slow and limited to fixed biological samples.

Here, we propose an alternative configuration to obtain directly Full Field SAF images (F$^2$-SAF) for a real time imaging of dynamic biological
processes. This approach has been introduced by Axelrod about ten years ago with a 1.65 NA objective lens~\cite{Axelrod2001}. Although the images showed a clear axial confinement, the lateral resolution was reduced by a factor 2. This certainly originates in the experimental set-up that did not permit accurate and quantitative measurements. In this paper, we give quantitative analysis of the optical characteristics of this technique and show that its performances should be reevaluated, thus opening the path to biological applications.

\begin{figure}
\centering\includegraphics[width=8.4cm]{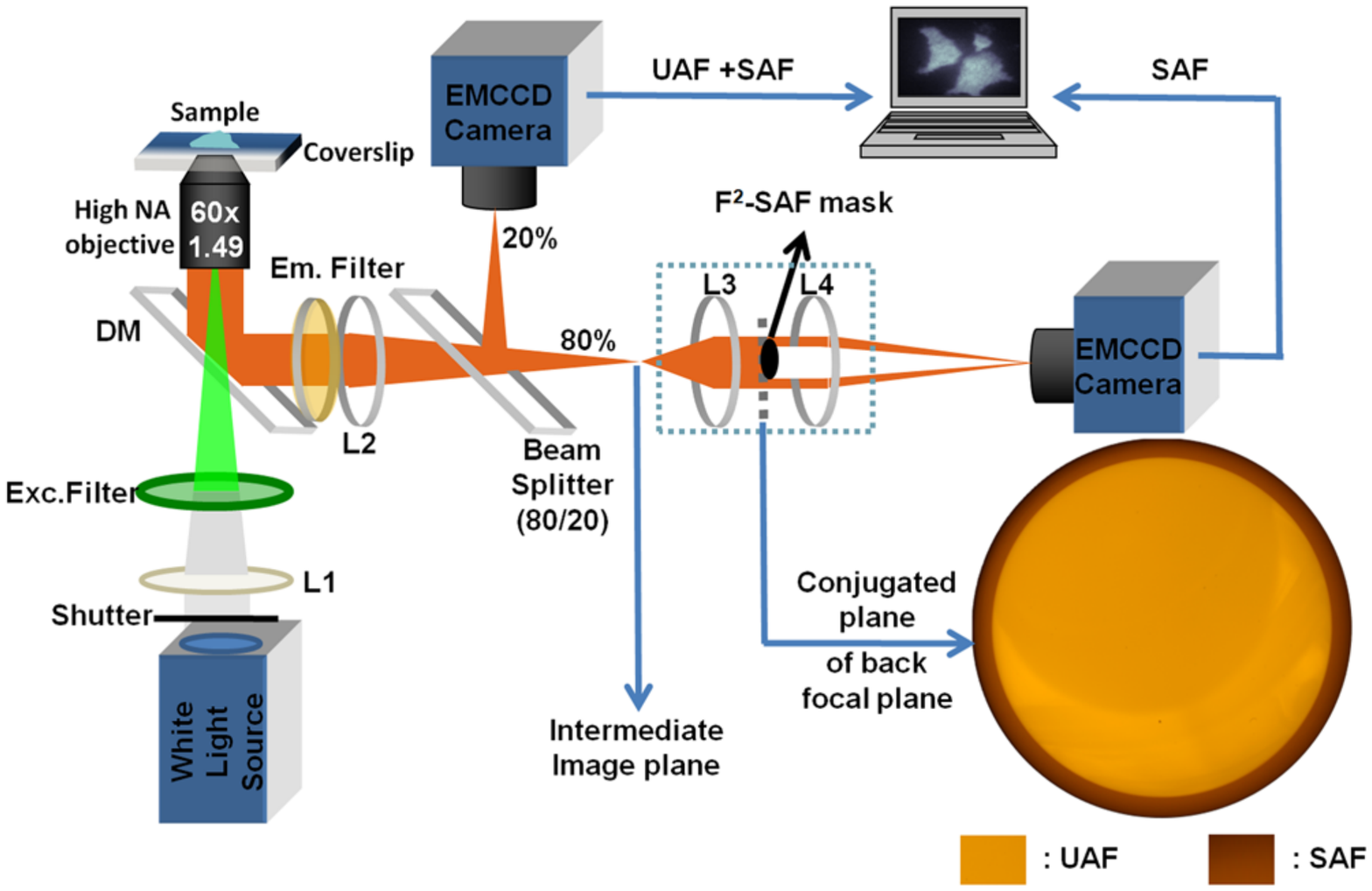}
\caption{Schematic of the experimental setup for multimodal real time imaging with two parallel channels for the epifluorescence and F$^2$-SAF
images.}
\label{setup}
\end{figure}

The schematic of the experimental setup is represented in figure~\ref{setup}. We use a commercial inverted microscope (Nikon Ti) with an
apochromat objective lens 1.49 NA 60x from Nikon and an EM-CCD camera (iXon+, Andor Technology). The illumination source is a standard fibred 130~W
mercury lamp which provides a homogeneous illumination on the sample. A two-lens system is included in the optical path to image the back focal
plane (BFP) of the objective on a conjugate plane. The relation between the radius $\rho$ in the BFP and the angle of emission $\theta_{em}$ is
given by the Abbe condition: $\rho(\theta_{em})=n_2 f\sin(\theta_{em})$, where $f$ is the focal distance of an apochromat objective and $n_2$ is
the refractive index of immersion medium. A circular mask in the shape of an opaque disk is placed on this conjugate plane to remove UAF. It is
mounted on a micrometric XYZ translation stage that allows fine positioning. An additional removable Bertrand lens can be added in the beam path to
image directly the BFP. This permits an easy centering and axial positioning of the mask. An image of the BFP is shown in figure~\ref{setup} for a
solution of Rhodamine (obtained with a Nikon camera Reflex D700). UAF and SAF areas are easily identifiable.

For biological studies, it can be of great interest to acquire in parallel the standard epifluorescence image (UAF+SAF), and the SAF image. In
order to obtain a simultaneous acquisition, a 20:80 beamsplitter cube has been added to split the beam into a UAF+SAF path (20\%) and a F$^2$-SAF
path (80\%) on two identical and synchronized EM-CCD cameras.

In the following paragraph, we focus on the optical performances of this F$^2$-SAF technique: the axial sectioning and the lateral resolution. We
use a method introduced by Mattheyses and Axelrod~\cite{Mattheyses2006} to measure the penetration depth. This technique takes advantage of the
intensity profile of 10~$\mu$m fluorescence beads to measure the penetration depth. The measured values are $160 \pm 15$~nm. This is in good
agreement with the Debye model. Besides, the penetration depth is constant all over the image.

The lateral resolution of the F$^2$-SAF is measured by acquiring the 2D Point Spread Function (PSF) of 100~nm fluorescent latex beads
(FluoSpheres$\textsuperscript{\textregistered}$ carboxylate-modified microspheres, Invitrogen). To reach a sufficient accuracy for the 2D PSF
measurements on the camera, we have added a 6.7x afocal optical system, composed of two doublet lenses of focal length $F_1=30$~mm and
$F_2=200$~mm, between the microscope and the camera. We thus obtain an over-sampling of 42~nm/pixel (far above the Nyquist criterion).

\begin{figure}
\centering\includegraphics[width=8.4cm]{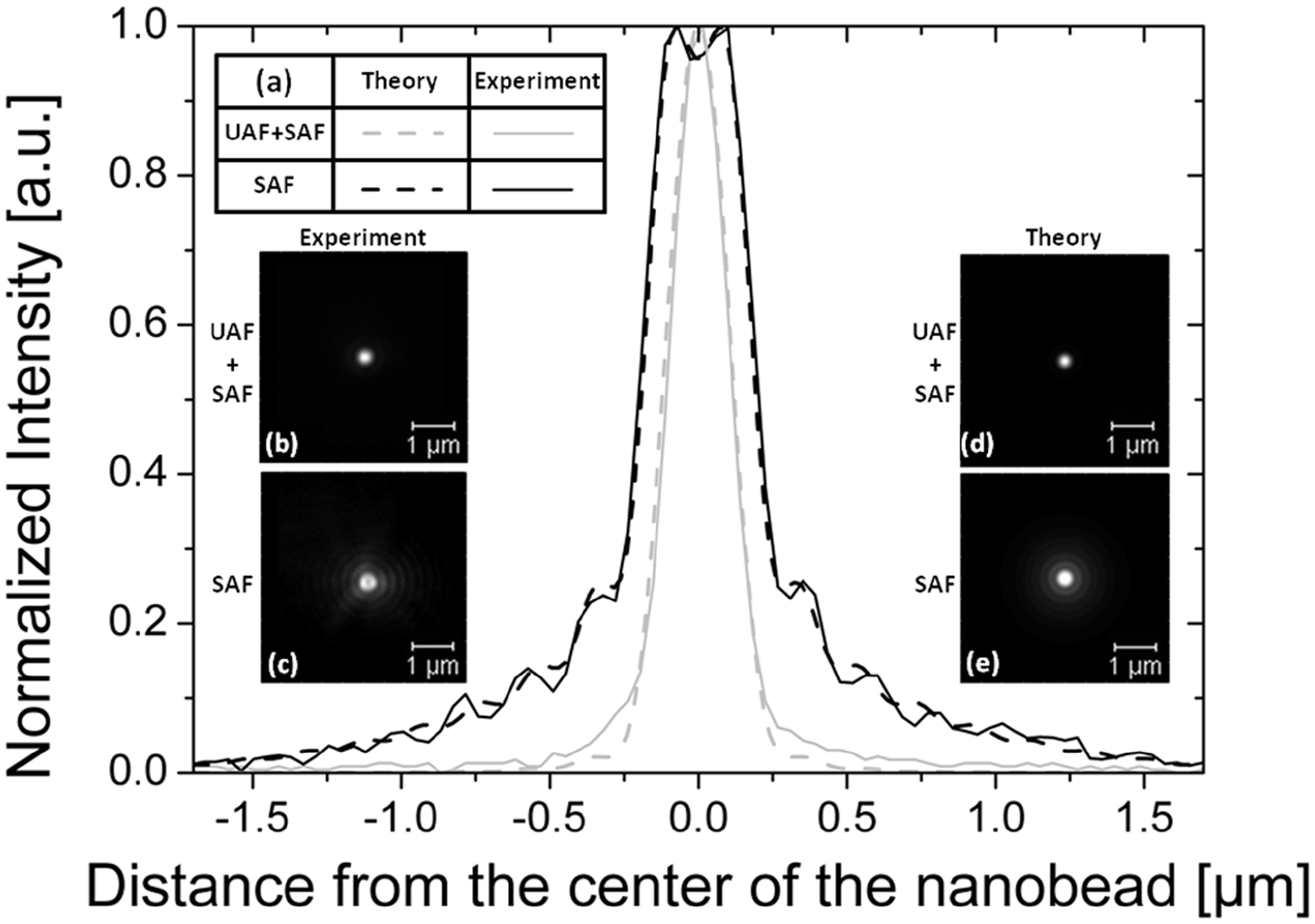}
\caption{(a) Normalized PSF profiles in UAF+SAF and F$^2$-SAF: model (dashed lines) and experiment (smooth lines). (b) and (c) show the PSF images
in UAF+SAF and F$^2$-SAF respectively, associated calculated images are shown in (d) and (e).}
\label{PSF}
\end{figure}

Figure~\ref{PSF}(a) shows the theoretical normalized cross-sectional profile (dashed lines) of the PSF obtained with the vectorial Debye integral
model which takes into account the polarization~\cite{Tang2007}. It is in good agreement with the experimental profiles (smooth lines) in UAF+SAF
and F$^2$-SAF. In the case of a dipole perpendicular to the surface, the emission is polarized radially with a minimum at the center. The dipole
average orientation has been used as a fitting parameter. A nearly isotropic distribution gives the best fit (with small enhanced orthogonal
proportion of 4\%). The images of a single nanobead are shown in figures~\ref{PSF}(b) for UAF+SAF and \ref{PSF}(c) for F$^2$-SAF. The associated
theoretical images are shown in figures~\ref{PSF}(d) and (e) respectively. As expected, we notice a loss of resolution on the F$^2$-SAF
PSF~\cite{Rivolta1986,Mahajan1986}. To evaluate the resolution, we compare the full width at half maximum (FWHM). The UAF+SAF PSF is known to be
fitted accurately by a gaussian\cite{Zhang2007}. The normalized intensity profile $I(r)$ is given by $I(r) = \exp( - \frac{r^2}{2\sigma^2})$ where
the standard deviation $\sigma$ is the only fitting parameter. The FWHM is given by: $W=~2\sqrt{2 \ln (2)}\sigma$. The gaussian curve fits
accurately the experimental data. For UAF+SAF collection, we obtain the value $W_{epi}=~198$~nm which is in good agreement with the one expected
from the theory: $W~=~196.9$~nm ($\sigma=~0.21\times \lambda_{em}/NA$). For SAF collection, we measure $W_{F^2-SAF}~=~247 \pm 5$~nm (average over
15 nanobeads). This represents a loss of resolution of about 25\% as compared to standard epifluorescence. We also pay a particular attention to
the study of any possible off-axis modifications of the PSF due to the abberations introduced by the objective lens. The value of $W$ has remained
consistent all over the image.

\begin{figure}
\centering\includegraphics[width=12cm]{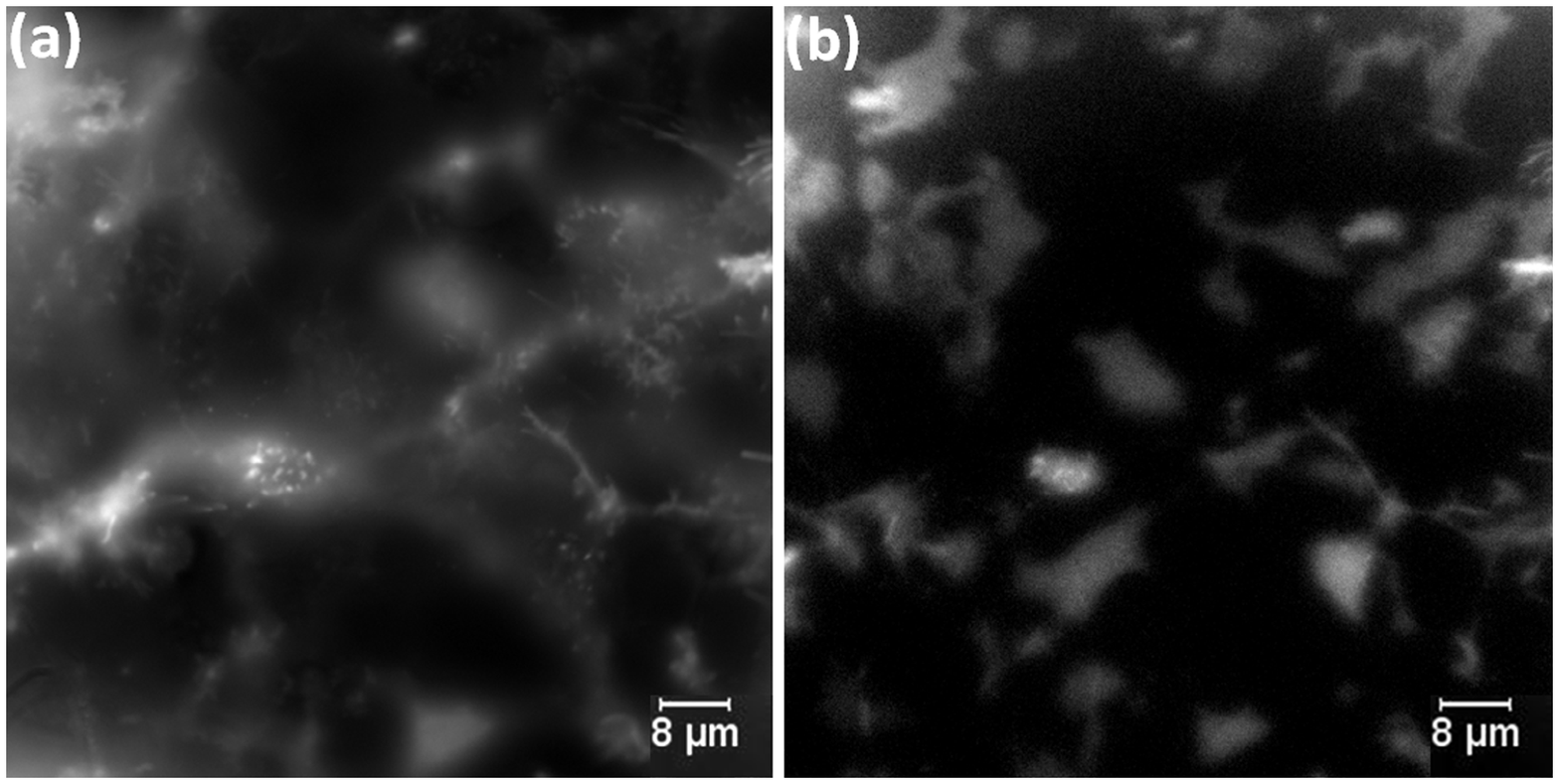}
\caption{HEK cells in (a) UAF+SAF and in (b) F$^2$-SAF.}
\label{HEK}
\end{figure}

We validate this full field imaging technique by performing real time cell imaging. We use Human Embryonic Kidney (HEK 293) cells transfected with type I cannabinoid receptors tagged with GFP (CB1R-GFP). Two cameras synchronized together with a time exposure of 300~ms have been used to obtain simultaneously images of the whole cell
and of its membrane. Figure~\ref{HEK}(a) shows HEK cells in UAF+SAF and their membranes \ref{HEK}(b) in F$^2$-SAF, where the fluorescence coming
from the inner parts of the cells is clearly removed. These images have been obtained with a simple white lamp. As expected, there is a slight loss
of lateral resolution in the F$^2$-SAF image, however the quality of the image is satisfactory. It is possible by using this configuration to
monitor membrane trafficking in real time.

In this paper, we have shown that F$^2$-SAF enables real time cell membrane imaging with a simple mask in the BFP of a commercial objective to
remove UAF components. We observed a slight loss in the lateral resolution which is due to the larger PSF of the fluorophores with emission dipoles
orthogonal to the surface. Nevertheless, the quality of the images permits us to easily resolve the structures of the membrane. Besides, image
quality could be further improved by using deconvolution techniques. The axial sectioning is of the order of 100~nm. Moreover, penetration depth
and lateral resolution performances are maintained throughout the whole image. F$^2$-SAF technique allows one to perform the observation of the
whole cell and its membrane simultaneously at low cost using a standard epifluorescence microscope and an incoherent source which provides an
homogeneous illumination all over the sample. We are currently applying this technique to biomedical issues, in particular the study of dictyostelium motility and endocytosis tracking. This technique could also find applications in other field like liquid-liquid interface studies. Ongoing developments focus on alternative optical configurations using removable masks with a single camera.

The authors thank C. Boccara and S. Gr\'{e}sillon for fruitful discussions, S. L\'{e}cart for cell samples, S. Sivankutty for helpful comments on
this article. This work has been supported by grants from Region Ile de France and C'Nano Ile de France.

\pagebreak

\section*{Informational Fourth Page}

\end{document}